# Experimental realization of single-plaquette gauge flux insertion and topological Wannier cycles

Zhi-Kang Lin[1,#], Ying Wu[2,#,†], Bin Jiang[1,#], Yang Liu[1], Shi-Qiao Wu[1], Feng Li[3,†], Jian-Hua Jiang[1,4,†]

[1]*School of Physical Science and Technology & Collaborative Innovation Center of Suzhou Nano Science and Technology, Soochow University, 1 Shizi Street, Suzhou, 215006, China*

[2]*School of Physics and Optoelectronics, South China University of Technology, 510640 Guangzhou, Guangdong, China*

[3]*Centre for quantum physics, Key laboratory of advanced optoelectronic quantum architecture and measurement (MOE), School of Physics, Beijing Institute of Technology, Beijing, 100081, China.*

[4]*Key Lab of Advanced Optical Manufacturing Technologies of Jiangsu Province & Key Lab of Modern Optical Technologies of Ministry of Education, Soochow University, Suzhou 215006, China.*

[#]These authors contributed equally to this work.

[†]Corresponding authors: jianhuajiang@suda.edu.cn (Jian-Hua Jiang), phlifeng@bit.edu.cn (Feng Li), phwuying@scut.edu.cn (Ying Wu).

## Abstract

**Gauge fields are at the heart of the fundamental science of our universe and various materials [1-4]. For instance, Laughlin's gedanken experiment of gauge flux insertion [2, 3] played a major role in understanding the quantum Hall effects. Gauge flux insertion into a single unit-cell, though crucial for detecting exotic quantum phases [4] and for the ultimate control of quantum dynamics and classical waves [5-7], however, has not yet been achieved in experiments. Here, we report on the realization of gauge flux insertion into a single plaquette in a lattice system with the gauge phase ranging from 0 to $2\pi$ which is achieved through a novel approach based on three consecutive procedures: the dimension extension, creating an engineered dislocation and the dimensional reduction. Furthermore, we discover that the single-plaquette gauge flux insertion leads to a new phenomenon termed as the topological Wannier cycles, i.e., the cyclic spectral flows**

**across multiple band gaps which are manifested as the topological boundary states (TBSs) on the plaquette. Such topological Wannier cycles emerge only if the Wannier centers are enclosed by the flux-carrying plaquette. Exploiting acoustic metamaterials and versatile pump-probe measurements, we observe the topological Wannier cycles by detecting the TBSs in various ways and confirm the single-plaquette gauge flux insertion by measuring the gauge phase accumulation on the plaquette. Our work unveils an unprecedented regime for lattice gauge systems and a fundamental topological response which could empower future studies on artificial gauge fields and topological materials.**

The Aharonov-Bohm effect [1] states that a system comes back to itself if a gauge flux of $2\pi$ is inserted (here, for simplicity, gauge flux is redefined as the gauge phase accumulated by going around the flux tube once anticlockwise), which, however, is true only for trivial systems. Inserting a gauge flux of $2\pi$ into a topological system leads to adiabatic pumping that yields spectral flows across the bulk band gap through topological boundary states, as first revealed in quantum Hall effects [2, 3] and later in other systems [4]. To date, local gauge flux insertion for electrons in solid state materials has been realized only in areas much larger than a unit-cell, as limited by the size of the magnetic flux tubes within the current technology [8-10]. Taking graphene as an example, a single-unit-cell gauge flux insertion can be realized only if a magnetic field larger than 20,000 Tesla is confined in the sub-nanometer scale, which is impossible within the current technology. For photons, phonons and ultracold atoms, although synthetic gauge fields have been realized in lattice systems using various methods [6, 7, 11-19], it is still challenging to realize gauge flux insertion in a single unit-cell, because such a regime requires extreme control of physical parameters in real-space.

Here, we utilize a novel approach to achieve synthetic gauge flux in a single plaquette in sonic crystals. Such extremely localized gauge flux is made possible due to the sharp structure configuration in the engineered screw dislocation as well as due to the dimension extension and dimensional reduction [20] procedures. In particular, the excellent controllability of sonic crystals fabricated by the three-dimensional (3D) printing technology gives access to sub-unit-cell structure engineering which enables the single-plaquette gauge flux. We remark that such controllability advantages over natural electronic materials may also be available in other synthetic materials and systems [6, 7, 21, 22].

Remarkably, the single-plaquette gauge flux insertion leads to topological Wannier cycles in topological crystalline insulators with filling anomaly which are featured by the mismatch between the Wannier centers and the unit-cell center [23-26] and are often regarded as higher-order topological insulators [24]. Materials with filling anomaly are found to be promising for catalytic, energy and other applications [25, 26]. Intriguingly, the topological Wannier cycles discovered here emerge only when the Wannier centers are enclosed by the flux-carrying plaquette. Therefore, the single-plaquette gauge flux can serve as an experimental probe of the Wannier centers with sub-unit-cell spatial resolution.

**Sonic crystals**

We use a two-dimensional (2D) sonic crystal (Fig. 1a) as the motherboard to demonstrate the single-plaquette gauge flux insertion and the topological Wannier cycles. The sonic crystal is designed as a square lattice of coupled acoustic resonators with the four-fold rotation ($C_4$) symmetry (lattice constant $a$=40 mm). In a unit-cell, there are four cylindrical resonators of the same diameter $D$=15mm and the same height $h$=25mm. The nearest-neighbor couplings are realized by the horizontal air tubes connecting the resonators. Specifically, the inter-cell (intra-cell) couplings are implemented by air tubes of the diameter $d_2 = 14$mm ($d_1 = 5$mm). The whole structure is encapsulated by photosensitive resin and is fabricated using the 3D printing technology.

The 2D sonic crystal realizes the 2D Su-Schrieffer-Heeger (SSH) model [27-29] which exhibits two bulk band gaps (denoted as gaps I and II, respectively; see Fig. 1b). These band gaps are acoustic analogs of higher-order topological insulators [24], as reflected by the fact that all the bands have their Wannier centers at the unit-cell corner (i.e., the Wyckoff position 1$b$; see Fig. 1c and the analysis in Supplementary Note 1). In this work, we denote such a case as topological, whereas the case with the Wannier centers at the unit-cell center (i.e., the Wyckoff position 1$a$) is denoted as trivial. The latter is realized by interchanging the inter-cell and intra-cell couplings in Fig. 1a.

**Synthetic gauge flux in a single plaquette**

To achieve the gauge flux confined in a single plaquette, we employ three consecutive procedures (see Fig. 1d): First, in the dimension extension procedure, we construct a 3D sonic crystal by stacking the 2D sonic crystals in Fig. 1a periodically along the $z$ direction. The interlayer couplings are realized by the vertical air tubes of the diameter $d=3$mm (Fig. 1e). The lattice constant along the $z$ direction is $H=36$ mm. Second, we create a step screw dislocation (SSD) at the center of the 3D sonic crystal with a Burgers vector $\boldsymbol{B}_v = (0, 0, H)$. The SSD divides the system into four quarter sectors that are related by the four-fold screw rotation symmetry $S_{4z} \coloneqq (x, y, z) \to (y, -x, z + \frac{H}{4})$. Each quarter sector is flat. The nearest-neighbor couplings across the boundary lines between adjacent quarter sectors are realized by the tilted tubes. After this procedure, the system becomes finite in the $x$ and $y$ direction but remains periodic in the $z$ direction. Third, via dimensional reduction [20], we map the 3D system into many $k_z$-dependent 2D systems by performing the Fourier transformation along the $z$ direction. After the transformation, $k_z$ becomes a parameter that drives the topological Wannier cycles in our system.

Since the interlayer couplings are rather weak, all $k_z$-dependent 2D systems have acoustic bands that are topologically equivalent to the original 2D acoustic bands in Fig. 1b. However, a key difference arises due to the SSD: The $k_z$-dependent 2D systems carry nontrivial gauge flux (Figs. 1e and 1f). In particular, after dimensional reduction, each tilted coupling between adjacent quarter sectors picks up a gauge phase of $\pm\frac{k_z H}{4}$ due to the $\pm\frac{H}{4}$ translation along the $z$ direction. This effect yields a hopping pattern depicted in Fig. 1f which indicates a gauge flux $\Phi = k_z H$ piercing through the central plaquette. Due to the periodicity in the $z$ direction, the gauge flux $\Phi = k_z H$ embraces the full phase range from 0 to $2\pi$.

**Topological Wannier cycles**

With the single-plaquette gauge flux insertion, eigenstates of different symmetries evolve cyclically between one and another within a group of four, as depicted in Fig. 2a. To illustrate this evolution, we label the eigenstates of a finite system by their $C_4$ eigenvalues $g_n = e^{in(\frac{2\pi}{4})}$ with $n = 0, \pm 1, 2$ corresponding to the $s$, $p_\pm = p_x \pm ip_y$, $d$-like states, separately. By inserting a flux quantum $\Phi = 2\pi$ into the central plaquette, an eigenstate with $C_4$ eigenvalue $g_n$ evolves into an eigenstate with $C_4$ eigenvalue $e^{i[n(\frac{2\pi}{4}) + \frac{\Phi}{4}]} = e^{i(n+1)(\frac{2\pi}{4})} = g_{n+1}$, because the gauge phase $\frac{\Phi}{4}$ is picked up upon each $C_4$ rotation (as elaborated in Supplementary Note 2). For the acoustic model in Fig. 1d, this scenario is kept because the $C_4$ rotation is replaced by the $S_{4z}$ screw rotation which acquires an additional phase $\frac{\Phi}{4}$ when acting on an eigenstate, due to the $\frac{H}{4}$ translation along the $z$ direction. We remark that although the system with an SSD breaks the chirality, the time-reversal symmetry remains intact. As a consequence, the gauge flux $\Phi = k_z H$ can vary either from 0 to $2\pi$ or from $2\pi$ to 0. In Fig. 2a, we only show the picture with $\Phi$ going from 0 to $2\pi$.

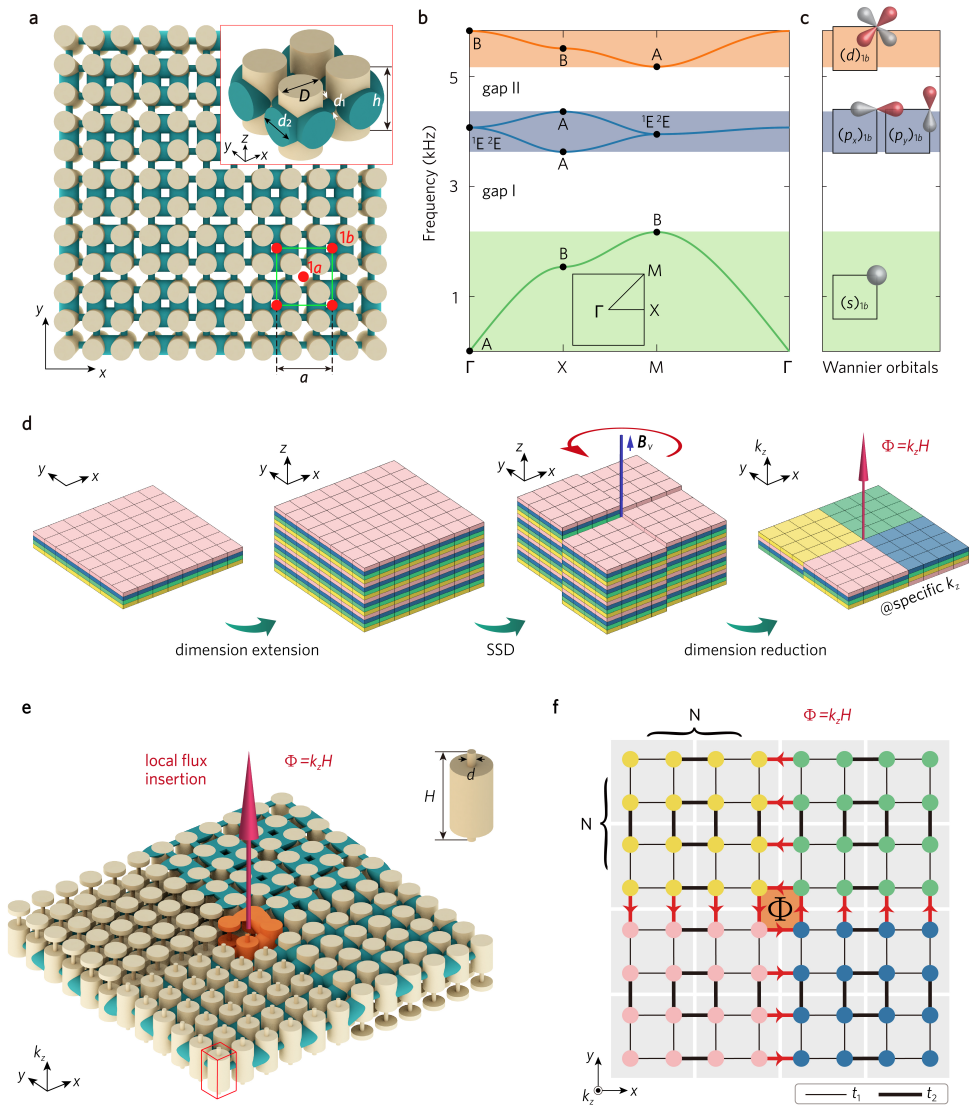

**Figure 1 | Synthetic gauge flux insertion in a single plaquette in acoustic systems**. **a**, A 2D sonic crystal mimicking the 2D SSH model. Yellow and dark-green structures depict the acoustic resonators and the tubes connecting these resonators, respectively. The green square depicts a unit-cell where the Wyckoff positions 1$a$ and 1$b$ are labeled. Inset shows the geometry details of a unit-cell. **b**, Acoustic

Bloch bands (colored regions) and topological band gaps (I and II). Symmetry properties of the bands are labeled by the little group representations at the high-symmetry points of the Brillouin zone (see Supplementary Note 1 for details). **c**, Illustration of the Wannier centers and Wannier orbitals of the Bloch bands. **d**, Schematic illustration of the procedures that realize the single-plaquette gauge flux. The blue line and arrow indicate the dislocation line and the Burgers vector, respectively. Four colors represent layers of different heights. **e,** The resultant structure when the procedures in **d** are applied to the sonic crystal in **a**. Inset shows that the interlayer couplings are realized by the tubes of a diameter $d$. The periodicity along the $z$ direction is $H$. Gauge flux insertion at the central plaquette (orange) is indicated. **f,** Tight-binding model corresponding to the acoustic structure in **e**. Each quarter sector has $N \times N$ unit-cells. The intra-cell and inter-cell couplings are denoted as $t_1$ and $t_2$, respectively. Interlayer coupling along the $z$ direction, $t_3$, is not shown. Red arrows denote the inter-sector couplings where a gauge phase of $\Phi/4$ is assigned for each hopping along the arrow.

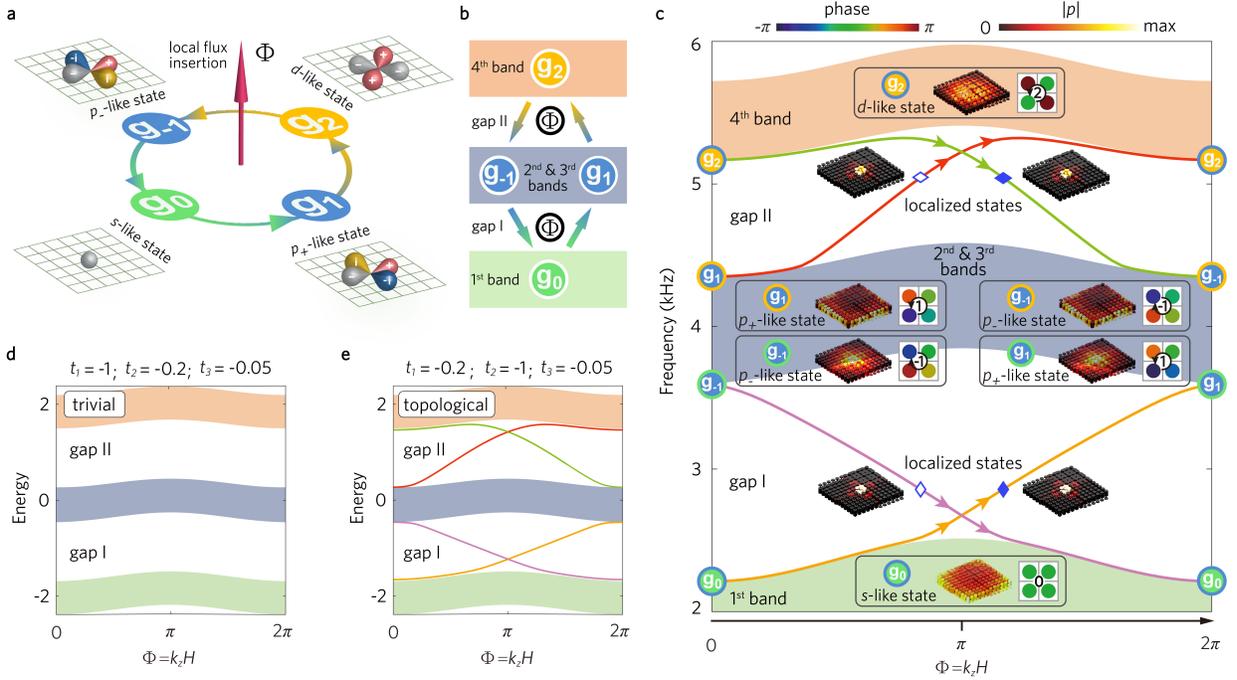

**Figure 2 | Topological Wannier cycles**. **a**, Schematic illustration of the evolution of eigenstates due to the single-plaquette gauge flux insertion. The eigenstates are labeled by their $C_4$ eigenvalues $g_n = e^{in(\frac{2\pi}{4})}$ with $n = 0, \pm 1, 2$ for the $s$, $p_\pm = p_x \pm i p_y$, $d$-like states, separately. The $C_4$-symmetry eigenvalues of the eigenstates evolve cyclically when $\Phi$ goes from 0 to $2\pi$. **b**, Schematic illustration of the cyclic spectral flows induced by the gauge flux insertion. **c**, Spectral flows traversing gaps I and

II as induced by gauge flux insertion confirmed by the simulation based on the acoustic model in Fig. 1e with 12×12 unit-cells in the *x-y* plane and periodic in the *z* direction. The bulk continua are painted with the same color scheme as in Fig. 1b. Eigenstates and their $C_4$-symmetry eigenvalues along the cyclic spectral flows are examined to reveal the evolution of the acoustic wavefunctions (including both the amplitude |*p*| and the phase of the acoustic pressure field as shown in the black boxes). The phase profiles are only shown for the four resonators at the system center. The phase winding directions are indicated by the black arrows, while the phase winding number is labeled at the center. For the in-gap localized states (labeled by the open and filled rhombuses), only the amplitude of the acoustic pressure field |*p*| is shown. **d-e**, Energy spectra of the bulk states (colored regions) and the TBSs (lines) of the tight-binding model in Fig. 1f for the trivial (**d**) and the topological (**e**) cases. The tight-binding parameters are listed on top of each figure. We use a supercell with 10×10 unit-cells in the *x-y* plane in the calculation.

Consider a system which is finite in the *x-y* plane, say, with $N^2$ unit-cells in each quarter sector as in Fig. 1f. If the system is a trivial atomic insulator, then each bulk band should have $4N^2$ states. In this case, the cyclic evolution of the eigenstates is within each bulk continuum and there is no spectral flow across the bulk band gaps. However, if the system is topological, the numbers of eigenstates in the bulk continua will deviate from the above. In our acoustic system, we find $4(N^2 \pm N) + 1$ eigenstates in the first and fourth bulk bands, respectively (see Supplementary Note 2 for more details). Meanwhile, the second and third bulk bands have $4(2N^2 - 1) + 2$ eigenstates in total. Note that these numbers are not integer multiples of four. Therefore, the cyclic evolution of eigenstates cannot be completed within each bulk continuum. There must be spectral flows across the bulk band gaps to fulfill the cyclic pattern in Fig. 2a when $\Phi$ goes from 0 to $2\pi$. Such spectral flows are termed as the topological Wannier cycles in this work.

The emergence of topological Wannier cycles can be interpreted through two alternative approaches. First, an intuitive way to understand the phenomenon is to adiabatically tune the system into a limit where the weak intra-cell couplings vanish. Such tuning keeps the band gaps open and the band topology unmodified, while the bulk states, on the other hand, become local Wannier orbitals confined in the plaquettes formed by the strong inter-cell couplings. In this limit, the single-plaquette gauge flux insertion affects only the Wannier orbitals confined in the central plaquette (Fig. 1f). As the central plaquette encloses the Wannier centers of the four acoustic bands, there are four Wannier orbitals in the central plaquette (one for each band, as shown in Fig. 1c). Thus, when a flux quantum $\Phi = 2\pi$ is gradually pierced into the central plaquette, the $s$-like Wannier orbital in the first band evolves into the $p_+$-like Wannier orbital in the second and third bands, meanwhile the original $p_+$-like Wannier orbital evolves into the $d$-like Wannier orbital in the fourth band. The spectral flows induced by the evolution of the Wannier orbitals is illustrated schematically in Fig. 2b. On the other hand, if the Wannier centers are not enclosed by the flux-carrying plaquette, the gauge flux will have no effect on the Wannier orbitals and there will be no spectral flow. The above analysis thus reveals that the topological Wannier cycles rely crucially on whether the Wannier centers are enclosed by the flux-carrying plaquette (see detailed analysis in Supplementary Note 3).

Alternatively, the topological Wannier cycles can be understood through the real-space topological invariants (RSTIs) [23, 30]. The imbalance of the $C_4$-symmetric eigenstates above and below a band gap is dictated by the RSTIs of the Bloch bands [30]. For gap I, the RSTIs are $\delta_1 = \delta_2 = -1$ from which we find that the imbalance of the $C_4$-symmetric eigenstates above and below gap I is predicted by $\Delta(g_0) = -\Delta(g_1) = \delta_1 = -1$, $\Delta(g_{-1}) = -\Delta(g_2) = \delta_2 - \delta_1 = 0$, for the process with $\Phi$ varying from 0 to $2\pi$ (see Supplementary Note 4 for details). Here, the $\pm$ signs stand for whether the excess state is above or below the band gap, respectively. Thus, there are one excess $s$-like state below gap I and one excess $p_+$-like state above gap I, which have to be connected by the spectral flow across gap I when $\Phi$ varies from 0 to $2\pi$. Simultaneously, there is a spectral flow which connects a $s$-like state below gap I with a $p_-$-like state above gap I, when $\Phi$ varies from $2\pi$ to 0, due to the time-reversal symmetry.

For gap II, we find that the RSTIs are $\delta_1 = 0, \delta_2 = -1$ when $\Phi$ varies from 0 to $2\pi$. The imbalance of the $C_4$-symmetric eigenstates above and below gap II is then predicted by $\Delta(g_0) = -\Delta(g_1) = \delta_1 = 0$, $\Delta(g_{-1}) = -\Delta(g_2) = \delta_2 - \delta_1 = -1$, for the process with $\Phi$ going from 0 to $2\pi$. Therefore, there are one excess $d$-like state above the gap and one excess $p_-$-like state below the gap which are connected by the spectral flow across gap II when $\Phi$ varies from 0 to $2\pi$. The time-reversal counterpart gives the spectral flow of a $d$-like state above gap II to a $p_+$-like state below the gap when $\Phi$ goes from $2\pi$ to 0. The above predictions from the RSTIs are consistent with the illustration in Fig. 2b and are further confirmed by the results in Fig. 2c which are obtained from the first-principle calculations based on the acoustic model in Fig. 1e (see Methods). Due to the gauge invariance, the spectral flows always start from a bulk band at $\Phi = 0$ and end at another bulk band at $\Phi = 2\pi$, since when $\Phi$ is an integer multiple of $2\pi$, there should be no TBS on the plaquette. Finally, the emergence (disappearance) of the topological Wannier cycles in the topological (trivial) phase is verified by the calculations based on the tight-binding model in Fig. 1f, as shown in Figs. 2d and 2e.

Interestingly, although both gaps I and II are acoustic analogs of higher-order topological insulators, they do not support any corner or edge state due to the chiral symmetry breaking, which is consistent with the recent theories [24, 31] and experiments [32]. Remarkably, here the topological Wannier cycles serve as an experimental probe of the higher-order topology even when the corner states fail to emerge. We emphasize that the above analysis unveils a new mechanism beyond the existing studies on the dislocation modes in topological insulators [33-44] (see Supplementary Notes 1-6 for details). In particular, the widely adopted theory in Ref. [33] cannot explain the findings in this work (see Supplementary Note 6).

**Experimental verification of topological Wannier cycles**

Before going into the experiments, we first show from calculations that the topological Wannier cycles emerge only in topological sonic crystals but vanish in trivial sonic crystals. This property is confirmed in Figs. 3a-b using the material parameters of the experimental sample. Here, the trivial sonic crystal is realized by interchanging the inter-cell and intra-cell couplings of the topological sonic crystal.

In experiments, a sonic crystal with an SSD (see Fig. 3c) is fabricated using the 3D printing technology based on the model in Fig. 1d. The sonic crystal is made of photosensitive resins which serve as the hard-wall boundaries for acoustic waves. Using a tiny microphone attached to a thin steel rod mounted on a translational stage, we can scan the 3D acoustic wavefunctions (i.e., the acoustic pressure fields) in the sample at various excitation frequencies (from 0.5kHz to 8kHz with a step of 9.4Hz). In all the measurements, an acoustic source is enclosed in either the top or the bottom resonator in hole 1 (denoted, separately, as the top or bottom excitation setups; see Methods). Through Fourier transformations of the detected acoustic pressure fields at each excitation frequency, we can extract the dispersions of the TBSs. For bottom (top) excitation, only the TBSs with positive (negative) group velocities are excited and measured.

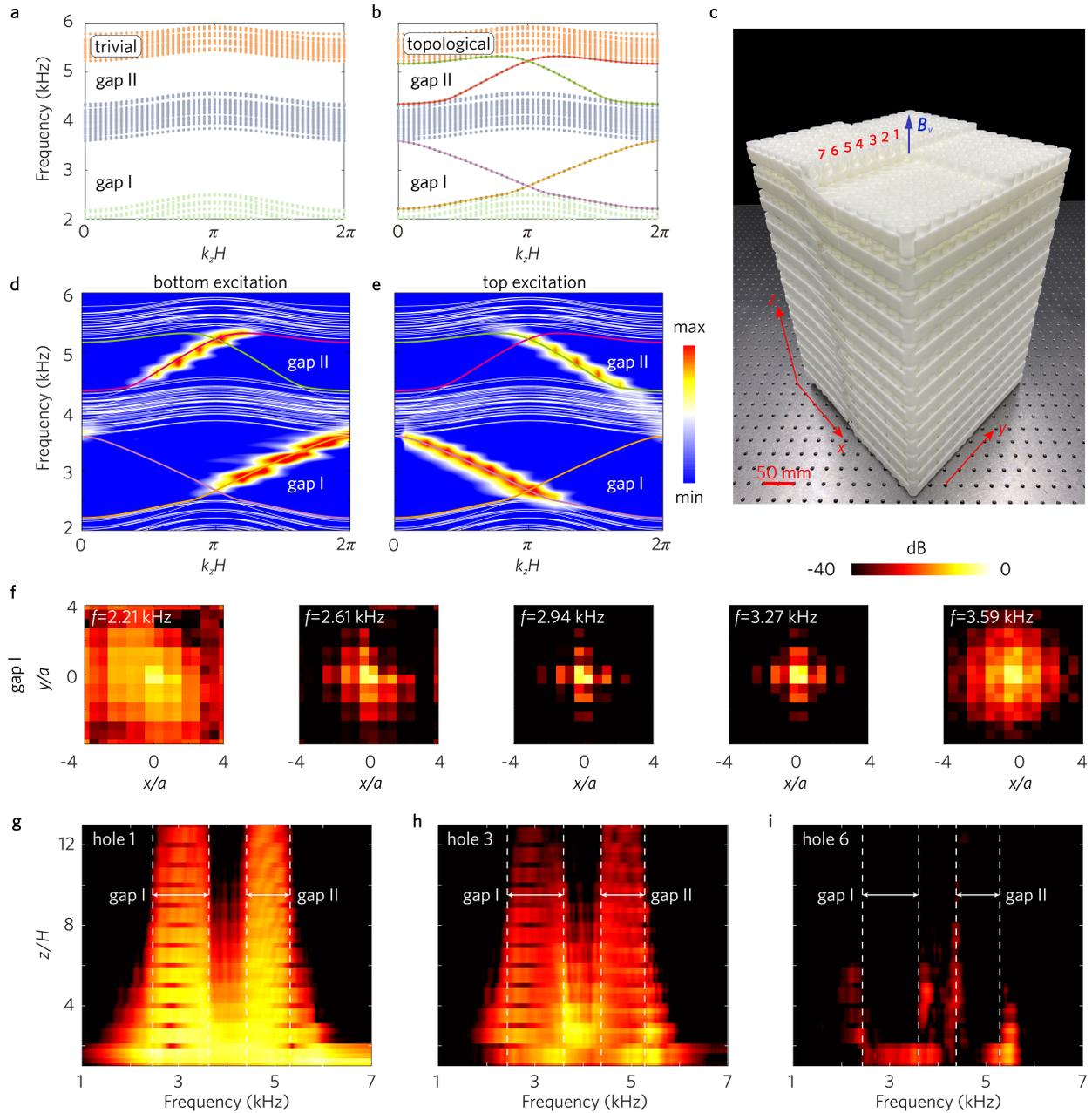

**Figure 3 | Observation of topological Wannier cycles**. **a**, Disappearance of the topological Wannier cycles in the trivial sonic crystal. **b**, Emergence of the topological Wannier cycles in the topological sonic crystal. The calculations in **a** and **b** are based on the acoustic model in Fig. 1d which is finite in the x-y plane with 8×8 unit-cells but periodic in the z direction. **c**, Photograph of the sonic crystal sample which has 13 layers in the z direction and 8×8 unit-cells in the x-y plane. The blue arrow

indicates the dislocation line and the Burgers vector. The numbers 1 to 7 label the holes through which we detect the acoustic wave propagations. **d-e**, Measured topological Wannier cycles in the setup with bottom and top excitations, respectively. **f,** Detected acoustic pressure profiles |p| in the *x-y* plane for five different excitation frequencies (labeled in each panel) across gap I in the bottom excitation setup. **g-i**, Detected acoustic pressure amplitude |p| in holes 1, 3 and 6, separately, as a function of the excitation frequency and the *z* coordinate of the detector in the bottom excitation setup.

As shown in Figs. 3d-e, the measured dispersions of the TBSs in both gaps I and II agree excellently with the calculation, except when the TBSs nearly merge into the bulk bands. In the latter case, the bulk states are also excited and their contributions to the detected acoustic signals cannot be filtered out. This mixing of signals from both the bulk states and the TBSs causes the deviation between the measured acoustic dispersions and the calculated spectrum of the TBSs. The unavoidable dissipation of acoustic waves and the finite-size effect of the sample broaden the measured dispersions in both frequency and wavevector dimensions, as analyzed in detail in Supplementary Notes 7 and 8. These effects, however, do not spoil the experimental observation of the topological Wannier cycles, as shown in Figs. 3d-e.

To further characterize the topological Wannier cycles, we present the measured acoustic pressure profiles in the *x-y* plane in Fig. 3f for five ascending excitation frequencies across gap I in the bottom excitation setup. The detection *x-y* plane is sufficiently away from (108 mm above) the acoustic source, to ensure that the measured acoustic pressure fields are mainly determined by the eigenstates wavefunctions, instead of the evanescent waves excited by the source. Figure 3f shows that as the excitation frequency gradually increases from below gap I to above gap I, the detected acoustic wavefunctions evolve from extended (bulk) states to localized states bound to the dislocation core (i.e., the TBSs), and then evolve again to extended (bulk) states. Such evolution of eigenstates is consistent with the scenario in Fig. 2c. Similar phenomena are observed across gap II (see Supplementary Note 7).

The spectral features can also be revealed by the transmission measurements in Figs. 3g-i. Results in Fig. 3g (for hole 1) show that the acoustic waves propagate persistently along the *z* direction in gaps I and II. In comparison, outside gaps I and II, acoustic waves decay rapidly in the *z* direction. Results in Fig. 3h (for hole 3) show similar features as in Fig. 3g. These observations indicate that the acoustic wave propagation near the core of the SSD is dominated by the modes in gaps I and II. In contrast, the results in Fig. 3i (for hole 6) exhibits opposite trends, indicating that the acoustic wave propagation away from the core of the SSD is mainly contributed by the modes outside gaps I and II. These experimental findings are consistent with the physical picture that, in gaps I and II, acoustic wave propagation is dominated by the TBSs bound to the core of the SSD, whereas outside these gaps, acoustic wave propagation is mainly through the extended bulk states (see Supplementary Note 7 for more details).

**Visualizing the TBSs and the single-plaquette gauge flux**

Here, we visualize experimentally the TBSs and characterize the gauge flux in the central plaquette quantitatively. For this purpose, we use the setup with the bottom excitation as depicted in Fig. 4a. In gaps I and II, the TBSs are localized around and propagate along the central plaquette (i.e., the core of the SSD) (Figs. 4a-b). By scanning the acoustic pressure field in the whole sample in the bottom excitation setup, we give the detected 3D acoustic wavefunction at the excitation frequency 3.0 kHz in Fig. 4b to visualize the TBSs directly. The acoustic structure in Fig. 4b is cut in a way to facilitate the 3D visualization of localized TBSs. It is seen that the TBSs are indeed localized around and propagate along the core of the SSD. Meanwhile, the localized acoustic wavefunctions around the core of the SSD in Fig. 4a, which are obtained from eigenstates calculation corresponding to the TBSs at 3.0 kHz, agree with the measured acoustic wavefunctions in Fig. 4b. The swirling of the acoustic energy flows in the inset of Fig. 4a indicates the possibly nonvanishing angular momentum of the TBSs. Numerical calculations confirm that the acoustic TBSs indeed carry notable angular momentum (see details in Supplementary Note 9).

The gauge flux inserted into the central plaquette can be calibrated by directly measuring the acoustic phase winding around the dislocation core. We focus particularly on the acoustic phases in the five resonators, A, B, C, D, and A', at the center of the sample (see the inset of Fig. 4b). From these phases, we can determine quantitatively the gauge phase accumulation when circulating the central plaquette once anticlockwise. In particular, the phase difference, $\varphi_{A'} - \varphi_A$, which is the gauge phase accumulated by circulating the central plaquette once anticlockwise, gives exactly the gauge flux inserted into the central plaquette $\Phi = k_z H$. In this way, the dependence of the gauge flux Φ on the excitation frequency should be consistent with the dispersion of the TBSs. As shown in Fig. 4c, the experimental results indeed show such consistency. Note that in the bottom excitation setup, only the branch of the TBSs with positive group velocities are excited in each band gap. The agreement between the calculated dispersion of the TBSs and the measured dependence of the gauge flux Φ on the excitation frequency are excellent, except when the group velocity of the TBSs becomes negative (and thus the TBSs become difficult to excite) or when the dispersion of the TBSs is close to the bulk continua. In the latter case, the simultaneously excited TBSs and bulk states mess up the measured phase difference and thus lead to the deviation between the measured gauge flux $\Phi = \varphi_{A'} - \varphi_A$ and the theoretical gauge flux $\Phi = k_z H$. More data and analysis are presented in Supplementary Note 10.

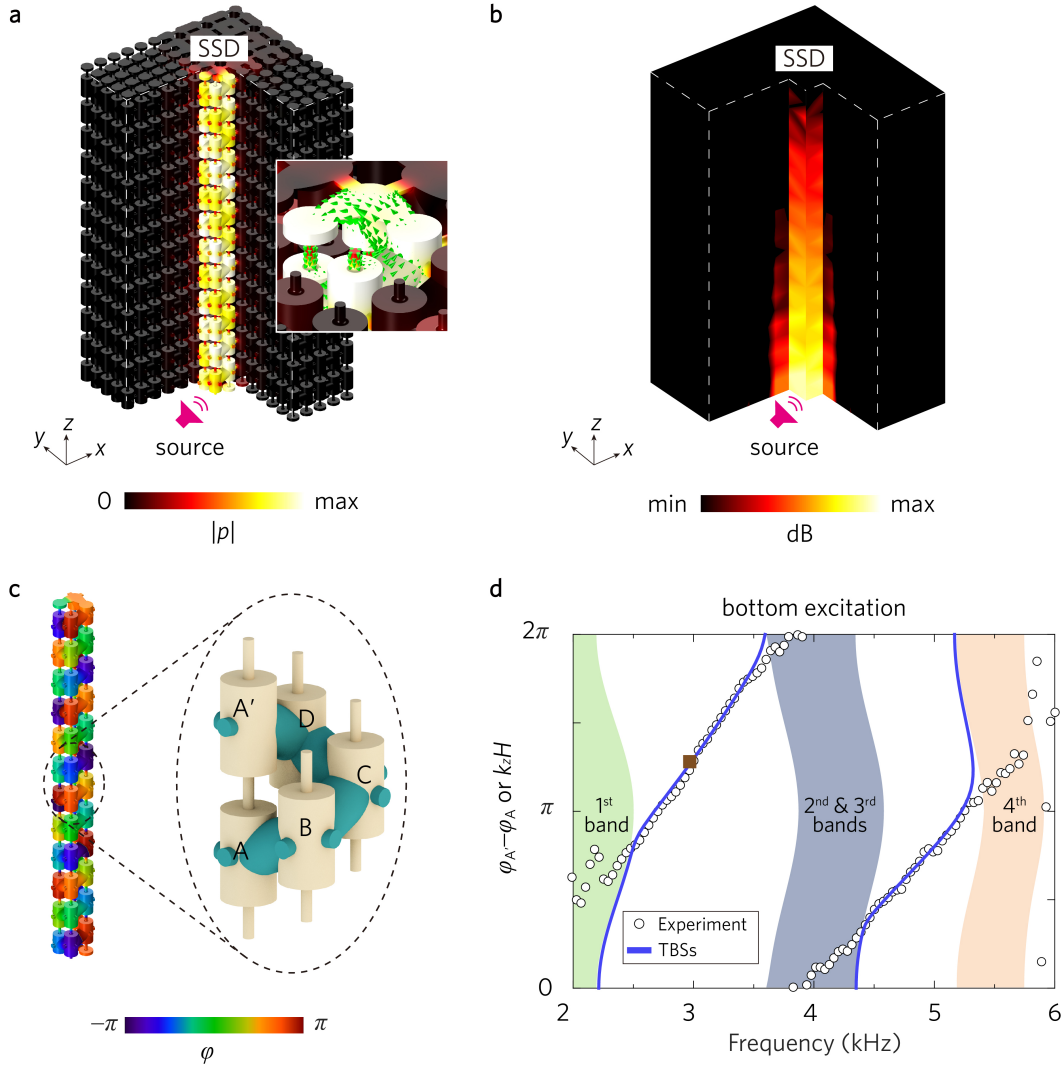

**Figure 4 | Visualizing the TBSs and the single-plaquette synthetic gauge flux**. **a**, Illustration of the TBSs localized around the central plaquette (i.e., the core of the SSD) as excited by an acoustic source in the bottom resonator in hole 1. Inset gives a zoom-in picture around the system center. Hot colors indicate the acoustic pressure amplitude |p|, while the green arrows indicate the distribution of the acoustic energy flows. **b**, The measured acoustic pressure profile |p| in the whole sample at the excitation frequency of 3.0 kHz (within gap I). **c,** Calculated 3D acoustic phase profile around the core of the SSD. Inset gives the zoom-in structure. **d**, The acoustic phase difference between the A' and A resonators, $\varphi_{A'} - \varphi_A$, as a function of the excitation frequency for the setup illustrated in **a**. The calculated spectrum of the acoustic system versus $k_z H$ is also presented in the figure for comparison: The blue curves give one branch of the TBSs in each band gap and the colored regions represent the bulk bands. The brown square labels the TBS excited and studied in **a-c**.

The direct measurements and visualization of the eigenstates in 3D space (especially the TBSs) and the transmission characterization of the system in Figs. 3 and 4, which have never been achieved in previous experimental studies of dislocations, demonstrate the advantages of the versatile measurements in acoustic systems.

**Discussions and outlook**

The realization of synthetic gauge flux in a single plaquette could inspire future studies in fundamental science and applications. For instance, such extremely localized gauge flux can be an invaluable tool for the engineering of novel topological states and the control of wave dynamics in both quantum and classical regimes [6, 7]. Our study unveils a new concept of topological Wannier cycles which are manifested as cyclic spectral flows in multiple band gaps. This phenomenon provides an unprecedented, powerful tool for the study of topological materials. In particular, it gives access to the probe of Wannier centers with sub-unit-cell spatial resolution which are crucial for the experimental investigation of various topological phases in crystalline materials and systems. Finally, our experiments also reveal a gap in the theory of topological defects in topological crystalline materials which is anticipated to be filled in the future.

# References

[1] Aharonov, Y. & Bohm, D. Significance of electromagnetic potentials in the quantum theory. *Phys. Rev.* **115**, 485–491 (1959).

[2] Laughlin, R. B. Quantized Hall conductivity in two dimensions. *Phys. Rev. B* **23**, 5632–5633 (R) (1981).

[3] Halperin, B. I. Quantized Hall conductance, current-carrying edge states, and the existence of extended states in a two-dimensional disordered potential. *Phys. Rev. B* **25**, 2185–2190 (1982).

[4] See, e.g., Wen, X. G. *Quantum Field Theory of Many-Body Systems*. Oxford Graduate Texts (Oxford University Press, Oxford, U.K, 2004).

[5] Nielsen, M. A. & Chuang, I. L. *Quantum Computation and Quantum Information* (Cambridge Univ. Press, 2000).

[6] Goldman, N., Juzeliūnas, G., Öhberg, P. & Spielman, I. B. Light-induced gauge fields for ultracold atoms. *Rep. Prog. Phys.* **77**, 126401 (2014).

[7] Aidelsburger, M. Nascimbene, S. & Goldman, N. Artificial gauge fields in materials and engineered systems. *C. R. Physique* **19**, 394-432 (2018).

[8] Tonomura, A. *et al.* Observation of Aharonov-Bohm effect by electron holography. *Phys. Rev. Lett.* **48**, 1443–1446 (1982).

[9] Albrecht, C., Smet, J. H., von Klitzing, K., Weiss, D., Umansky, V. & Schweizer, H. Evidence of Hofstadter's fractal energy spectrum in the quantized Hall conductance. *Phys. Rev. Lett.* **86**, 147–150 (2001).

[10] Geim, A. K., Bending, S. J. & Grigorieva, I. V. Asymmetric scattering and diffraction of two-dimensional electrons at quantized tubes of magnetic flux. *Phys. Rev. Lett.* **69**, 2252–2255 (1992).

[11] Fang, K., Yu, Z. & Fan, S. Realizing effective magnetic field for photons by controlling the phase of dynamic modulation. *Nat. Photon.* **6**, 782–787 (2012).

[12] Rechtsman, M. C. *et al.* Strain-induced pseudomagnetic field and photonic landau levels in dielectric structures. *Nat. Photon.* **7**, 153–158 (2013).

[13] Mittal, S., Ganeshan, S., Fan, J., Vaezi, A. & Hafezi, M. Measurement of topological invariants in a 2D photonic system. *Nat. Photon.* **10**, 180–183 (2016).

[14] Jia, H. *et al.* Observation of chiral zero mode in inhomogeneous three-dimensional Weyl metamaterials. *Science* **363**, 148–151 (2019).

[15] Lumer, Y. *et al.* Light guiding by artificial gauge fields. *Nat. Photon.* **13**, 339–345 (2019).

[16] Xiao, M., Chen, W.-J., He, W.-Y. & Chan, C. T. Synthetic gauge flux and Weyl points in acoustic systems. *Nat. Phys.* **11**, 920-924 (2015).

[17] Wen, X., Qiu, C., Qi, Y., Ye, L., Ke, K., Zhang, F. & Liu, Z. Acoustic Landau quantization and quantum-Hall-like edge states. *Nat. Phys.* **15**, 352–356 (2019).

[18] Peri, V., Serra-Garcia, M., Ilan, R. & Huber, S. D. Axial-field-induced chiral channels in an acoustic Weyl system. *Nat. Phys.* **15**, 357–361 (2019).

[19] Fang, K. *et al.* Generalized non-reciprocity in an optomechanical circuit via synthetic magnetism and reservoir engineering. *Nat. Phys.* **13**, 465–471 (2017).

[20] Qi, X.-L., Hughes, T. L. & Zhang, S.-C. Topological field theory of time-reversal invariant insulators. *Phys. Rev. B* **78**, 195424 (2008).

[21] Ma, G., Xiao, M. & Chan, C. T. Topological phases in acoustic and mechanical systems. *Nat. Rev. Phys.* **1**, 281–294 (2019).

[22] Ozawa, T. *et al.* Topological photonics. *Rev. Mod. Phys.* **91**, 015006 (2019).

[23] van Miert, G. & Ortix, C. Higher-order topological insulators protected by inversion and rotoinversion symmetries. *Phys. Rev. B* **98**, 081110(R) (2018).

[24] Benalcazar, W. A., Li, T. & Hughes, T. L. Quantization of fractional corner charge in $C_n$-symmetric higher-order topological crystalline insulators. *Phys. Rev. B* **99**, 245151 (2019).

[25] Gao, J. *et al.* Unconventional materials: the mismatch between electronic charge centers and atomic positions. Preprint at https://arXiv.org/abs/2106.08035

[26] Xu, Y. *et al.* Filling-enforced obstructed atomic insulators. Preprint at https://arXiv.org/abs/2106. 10276

[27] Liu, F. & Wakabayashi, K. Novel topological phase with a zero Berry curvature. *Phys. Rev. Lett.* **118**, 076803 (2017).

[28] Xie, B.-Y. *et al.* Second-order photonic topological insulator with corner states. *Phys. Rev. B* **98**, 205147 (2018).

[29] Xie, B.-Y. *et al.* Higher-order band topology. Nat. Rev. Phys. **3**, 520–532 (2021).

[30] Song, Z.-D., Elcoro, L. & Bernevig, B. A. Twisted bulk-boundary correspondence of fragile topology. *Science* **367**, 794–797 (2020).

[31] Peterson, C. W. *et al.* A fractional corner anomaly reveals higher-order topology. *Science* **368**, 1114–1118 (2020).

[32] van Miert, G. & Ortix, C. On the topological immunity of corner states in two-dimensional crystalline insulators. *npj Quantum Materials* **5**, 63 (2020).

[33] Ran, Y., Zhang, Y. & Vishwanath, A. One-dimensional topologically protected modes in topological insulators with lattice dislocations. *Nat. Phys.* **5**, 298-303 (2009).

[34] Teo, J. C. Y. & Kane, C. L. Topological defects and gapless modes in insulators and superconductors. *Phys. Rev. B* **82**, 115120 (2010).

[35] Juričić, V., Mesaros, A., Slager, R.-J. & Zaanen, J. Universal probes of two-dimensional topological insulators: dislocation and π-flux. *Phys. Rev. Lett.* **108**, 106403 (2012).

[36] de Juan, F., Rüegg, A. & Lee, D.-H. Bulk-defect correspondence in particle-hole symmetric insulators and semimetals. *Phys. Rev. B* **89**, 161117(R) (2014).

[37] Slager, R.-J., Mesaros, A., Juričić, V. & Zaanen, J. Interplay between electronic topology and crystal symmetry: dislocation-line modes in topological band insulators. *Phys. Rev. B* **90**, 241403(R) (2014).

[38] van Miert, G. & Ortix, C. Dislocation charges reveal two-dimensional topological crystalline invariants. *Phys. Rev. B* **97**, 201111(R) (2018).

[39] Queiroz, R., Fulga, I. C., Avraham, N., Beidenkopf, H. & Cano, J. Partial lattice defects in higher-order topological insulators. *Phys. Rev. Lett.* **123**, 266802 (2019).

[40] Slager, R.-J. The translational side of topological band insulators. *J. Phys. Chem. Solids* **128**, 24-38 (2019).

[41] Roy, B. & Juričić, V. Dislocation as a bulk probe of higher-order topological insulators. Preprint at https://arXiv.org/abs/2006.04817

[42] Paulose, J., Chen, B. G. & Vitelli, V. Topological modes bound to dislocations in mechanical metamaterials. *Nat. Phys.* **11**, 153-156 (2015).

[43] Li, F.-F. *et al.* Topological light-trapping on a dislocation. *Nat. Commun.* **9**, 2462 (2018).

[44] Nayak, A. K. *et al.* Resolving the topological classification of bismuth with topological defects. *Sci. Adv.* **5**, eaax6996 (2019).

## Methods

### Simulations

We performed systematic finite-element simulations for the acoustic waves in the 3D printed sonic crystal structures using COMSOL Multiphysics with the pressure acoustic module. Due to the huge acoustic impedance mismatch between air and the photosensitive resin used in the 3D printing, the latter can be treated as sound hard boundaries in the simulation. Sound waves propagate in air with a mass density of 1.25 kg/m$^3$ at the speed of 343 m/s at room temperature (23 °C). The bulk band dispersions of the acoustic waves in the 2D sonic crystal (Fig. 1b in the main text) are calculated using a single unit cell (shown in the inset of Fig. 1a in the main text) with Floquet Bloch boundary conditions in both the $x$- and $y$-directions. On the other hand, to calculate the eigenstates of the structure with a SSD, we treat the system as periodic in the $z$-direction but finite in the $x$- and $y$- directions. Specifically, using the supercell shown in Fig. 1d in the main text, we calculate the acoustic dispersions with the Floquet Bloch boundary condition in the *z*-direction and the closed boundary conditions in the *x*- and *y*- directions. Simultaneously, the acoustic wavefunctions (i.e., the amplitude and phase profiles of the acoustic pressure and velocity fields) of the eigenstates are obtained.

**Experiments**

The sample was manufactured by 3D printing technology using photosensitive resin and was assembled through layer-by-layer stacking up to 13 layers. To measure the TBSs, a headphone of a diameter 6mm is utilized for acoustic excitations with the frequency sweeping from 0.5kHz to 8kHz at a step of 9.4Hz. The headphone is placed and enclosed in either the top or the bottom resonator of hole 1 in the sample (denoted as the top or bottom excitation, respectively). A tiny microphone (2.5mm×1.1mm×3.3mm) mounted on a steel rod is connected with the network analyzer (Keysight E5061B) and inserted into the sample to detect the acoustic pressure profile (see Supplementary Note 7). Automatic scanning with 9mm steps in the *z*-direction in each hole as driven by a translational stage can be used to measure the acoustic pressure profiles in the whole sample, if needed. Such measurements contain both the amplitude and phase profiles of the acoustic pressure field, thanks to the data processing by the network analyzer. Through fast Fourier transformations of the detected acoustic pressure profiles at each excitation frequency, we can extract the dispersions of the acoustic waves in the sample. Specifically, this method is utilized to determine experimentally the dispersions of the TBSs. Such acoustic pump-probe measurements are also used to analyze the phase accumulations around the central plaquette (see Supplementary Note 10). To directly visualize the acoustic wavefunctions of the TBSs inside the sample (as shown in Fig. 4b), we measure the acoustic pressure profiles in the whole sample in the bottom excitation setup.

**Acknowledgements**

Z.K.L, B.J, Y.L, S.Q.W and J.H.J are supported by the National Natural Science Foundation of China (Grant No. 12074281), the Jiangsu Province Distinguished Professor Funding and the Key Lab of Advanced Optical Manufacturing Technologies of Jiangsu Province, Soochow University. J.H.J thanks Dr. Zhi-Da Song for many helpful discussions. He also thanks Dr. Zhi Hong Hang for sharing his laboratory for part of the measurements and the Huazhong University of Science and Technology for hospitality where a large part of the manuscript was finalized. Y.W and F.L are supported by the Natural Science Foundation of Guangdong Province (Grant No. 2020A1515010549) and China Postdoctoral Science Foundation (Grant No. 2020M672615).

## Author contributions

J.H.J initiated the project and guided the research. J.H.J and Z.K.L established the theory. Y.W, Z.K.L, B.J and Y.L performed the numerical calculations and simulations. Y.W, Z.K.L, S.Q.W, J.H.J and F.L designed and performed the experiments. All the authors contributed to the discussions of the results and the manuscript preparation. J.H.J, Z.K.L and Y.W wrote the manuscript and the Supplementary Information.

## Competing Interests

The authors declare that they have no competing financial interests.

## Data availability

All data are available in the manuscript and the Supplementary Information. Additional information is available from the corresponding authors through reasonable request.

## Code availability

We use the commercial software COMSOL MULTIPHYSICS to perform the acoustic wave simulations and eigenstates calculations. Reasonable request to computation details can be addressed to the corresponding authors.